# Probabilistic Visual Secret Sharing Schemes for Gray-scale images and Color images[*]

Dao-Shun Wang, Feng Yi and Xiaobo Li, Senior Member, IEEE

Abstract— Visual secrete sharing (VSS) is an encryption technique that utilizes human visual system in the recovering of the secret image and it does not require any complex calculation. Pixel expansion has been a major issue of VSS schemes. A number of probabilistic VSS schemes with minimum pixel expansion have been proposed for binary secret images. This paper presents a general probabilistic ($k$, $n$)-VSS scheme for gray-scale images and another scheme for color images. With our schemes, the pixel expansion can be set to a user-defined value. When this value is 1, there is no pixel expansion at all. The quality of reconstructed secret images, measured by Average Relative Difference, is equivalent to Relative Difference of existing deterministic schemes. Previous probabilistic VSS schemes for black-and-white images with respect to pixel expansion can be viewed as special cases of the schemes proposed here

## Ⅰ. INTRODUCTION

Visual secret sharing (VSS) schemes have been proposed to encode a secret image into $n$ "shadow" images ("shares"). In a ($k$, $n$)-VSS scheme, the secret can be visually reconstructed only when $k$ or more shares are available. Each pixel of the secret image is "expanded" into

[*] This research was supported in part by the National Natural Science Foundation of China under Grant No. 90304014, Basic Research Foundation of School of Information Science and Technology of Tsinghua and Canadian Natural Sciences and Engineering Research Council under Grant OGP9198.

The authors are with the Department of Computer Science and Technology, Tsinghua University, Beijing, 100084, China, and Department of Computing Science, University of Alberta, Edmonton, Alberta, T6G 2E8,Canada. (e-mail: daoshun@mail.tsinghua.edu.cn, li@cs.ualberta.ca).

$m$ sub-pixels in each share. In the reconstruction process, the stacking of the sub-pixels is a Boolean "OR" operation. VSS schemes were primarily designed for black-and-white (binary) images [1] by Naor and Shamir. Based on the definition of [1], Verheul and Van Tilborg [3] gave a more general definition. Directly based on black-and-white schemes, VSS schemes for gray-scale images (called GVSS) with pixel expansion $m^*$ are proposed [2, 3, 4], and VSS for color images (called CVSS) with pixel expansion $m'$ are proposed [3, 5]. All those schemes are "deterministic" since every pixel in the secret image can be exactly reconstructed, but each share is $m$ times bigger than the original. For example, when a (3, 3)-GVSS scheme proposed in [2] is used to code an image with 256 grey-levels, $m^*$ is as big as 1020. This heavy space requirement makes it impractical in many applications. To solve this pixel expansion problem, Ito et al. [6] and Yang [7] proposed probabilistic VSS (Prob. VSS) schemes for binary images with no pixel expansion ($m = 1$). Small areas, instead of individual pixels, of the secret image can be correctly reconstructed. The size of the recognized area is analyzed in [7]. Cimato et al. [8] extended the probabilistic binary VSS model that allows $m \geq 1$ and discussed the relation between probabilistic schemes and deterministic schemes. Hsu et al. [9] used the concept of probability to construct an optimization model for general access structures. So far, all existing probabilistic VSS schemes [6, 7, 8, 9] are for binary images.

In this paper, we propose two probabilistic VSS schemes, one for gray-scale images and one for color images. As in [8], the pixel expansion can be set for a particular value $m \geq 1$, and the size of a recognizable area in [7] is analyzed in detail. Table I lists the proposed schemes and related probabilistic schemes. In this table and the rest of the paper, we include the pixel expansion value and the secret image type into the name of each scheme. In this notation, $g$ is the number of distinct grey-levels in the secret image, $c$ is the number of distinct colors in the secret image. The parameters $m$, $m^*$ and $m'$ are the pixel expansion the for basic binary VSS, gray-scale image GVSS and color image CVSS, respectively. The quality of the reconstructed secret image is measured by the "contrast" that is the relative difference between consecutive grey-levels. The detailed definition of this measure is given in sub-section E of the Section Ⅱ. Section Ⅱ of this paper gives the background and basic notation of previous GVSS and CVSS directly based on the black-and-white VSS schemes, and defines the quality measure relative difference. Section III presents our probabilistic (*k*, *n*, *m*\*, *g*)-GVSS scheme. The analysis on the size of the recognized area for this scheme is in Section IV. In Section V, we introduce a method to construct color probabilistic (*k*, *n*, *m'*, *c*)-CVSS scheme. The size of the recognized area for color probabilistic CVSS scheme is presented in section VI. Finally, conclusions are given in Section VII.

TABLE I

THE PROPOSED (*k*, *n*)-VSS SCHEMES AND RELATED PROBABILISTIC SCHEMES

|  | Prob. VSS in [6] | Prob. VSS in [7] | Prob. GVSS proposed here | Prob. CVSS proposed here |
|---|---|---|---|---|
| Image type | binary | Binary | gray-scale | Color |
| Pixel expansion | 1 | 1, …, *m* | 1, …, *m*\* | 1, …, $m'$ |
| Region size | analyzed | not analyzed | analyzed | Analyzed |
| Relative difference | $\alpha$ | give computation equations | $\alpha_i, i=0,\cdots,g-2$ | $\alpha'_j, j=0,\cdots,c-1$ |

The value $\alpha$ is the relative difference of deterministic black–and-white VSS schemes. The value $\alpha_i$ is the relative difference of deterministic GVSS schemes for gray-scale images. The value $\alpha'_j$ is the relative difference of deterministic color CVSS schemes.

## II. BACKGROUND AND BASIC NOTATIONS

### A. Binary $(k, n)$-VSS schemes

In black–and-white VSS, the secret image consists of a collection of black–and-white pixels and each pixel is subdivided into a collection of *m* black–and-white sub-pixels in each of the *n* shares. The collection of sub-pixels can be represented by an $n \times m$ Boolean matrix *S* = [$s_{ij}$], where the element $s_{ij}$ represents the *j-th* sub-pixel in the *i-th* share. A white sub-pixel is represented as a 0, and a black sub-pixel is represented as a 1. The white sub-pixels let through the light while black sub-pixels stop it. $s_{ij} = 1$ if and only if the *j-th* sub-pixel in the *i-th* share is black. The grey-level of the combined share obtained, by stacking shares $i_1, \ldots, i_r$ is proportional to the Hamming weight (the number of 1's in the vector $V$) $H(V)$ of the OR-ed ("OR" operation) $m$-vector $V = OR(i_1, \cdots, i_r)$ where $i_1, \ldots, i_r$ are the rows of $S$ associated with the shares we stack. Verheul and Van Tilborg [3] extended the definition of Naor and Shamir's scheme [1]. Let $z(V)$ denote the number of zero coordinates of a vector $V$. Note that $H(V) + z(V) = m$, the $H(V)$ is the Hamming weight of the vector $V$. This grey-level is interpreted as black by the user's visual system if $H(V) \leq l$, and as white if $H(V) \geq h$ ($h > l$). In [1], the contrast between combined shares that come from a white pixel and a black pixel is implicitly defined as $h - l$, and the loss of the contrast in [3] as $\alpha = (h-l)/m$, namely the relative difference $\alpha = \dfrac{h-l}{m}$.

Example 1 below demonstrate Definition [3].

***Example 1:*** A (2, 3, 3)-VSS scheme for black–and-white image.

The basic matrices $S^0$ and $S^1$ for black–and-white (2, 3, 3)-VSS scheme ([1]).

$$S^0 = \begin{bmatrix} 001 \\ 001 \\ 001 \end{bmatrix} \qquad S^1 = \begin{bmatrix} 100 \\ 010 \\ 001 \end{bmatrix}$$

The OR of any two rows in $S^0$ has two the number of zeros, namely $h = 2$ and the OR of any two rows in $S^1$ has one the number of zeros, here $l = 1$. Thus, the relative difference between a reconstructed black pixel and a reconstructed white pixel is $\alpha = \dfrac{h-l}{m} = 1/3$.

**B. Gray-scale $(k, n, m^*, g)$-GVSS schemes**

The following definition of a $k$-out-of-$n$ secret sharing scheme for grey-level images is from [2].

***Definition 1[2, 4]:*** Let $g \geq 2$ be an integer. The $g$ collections of $n \times m^*$ basis matrices $G^0, \cdots, G^{g-1}$ construct a $k$-out-of-$n$ visual secret sharing scheme for $g$ grey-levels with pixel expansion $m^*$ ($(k, n, m^*, g)$-GVSS Scheme, for short). A $(k, n, m^*, g)$-GVSS Scheme with relative difference $\alpha_0, \cdots, \alpha_{g-2}$ and sets of thresholds $\{d_i\}$, $i = 0, \cdots, g-2$, is realized using $n \times m$ basis matrices $G^0, \cdots, G^{g-1}$ if the following two conditions hold.

1. For $i = 0, \cdots, g-2$, the Hamming weight of the "OR" $m^*$-vector $V^i$ of any of $k$-out-of-$n$ row in $G^i$ satisfies $H(V^i) \leq d_i - \alpha_i \cdot m^*$; whereas, for $G^{i+1}$ it results that $H(V^{i+1}) \geq d_i$.

2. For any subset $\{r_1, \cdots, r_j\} \subset \{1, \cdots, k\}$ with $j < k$, the $g$ $j \times m^*$ matrices obtained by restricting $G^0, \cdots, G^{g-1}$ to rows $r_1, \cdots, r_j$ are equal up to a column permutation.

The following example shows the construction method of a gray-scale $(k, n, m^*, g)$-GVSS Schemes from a binary $(k, n, m)$-VSS scheme.

***Example 2:*** *(continuation of Example 1)*:

The matrices $G^0$, $G^1$, and $G^2$ are the basis matrices of a (2, 3, 6, 3)-GVSS scheme. They are constructed as follows.

$$G^0 = S^0 \circ S^0 = \begin{bmatrix} 001 \\ 001 \\ 001 \end{bmatrix} \circ \begin{bmatrix} 001 \\ 001 \\ 001 \end{bmatrix} = \begin{bmatrix} 001001 \\ 001001 \\ 001001 \end{bmatrix}$$

$$G^1 = S^0 \circ S^1 = \begin{bmatrix} 001 \\ 001 \\ 001 \end{bmatrix} \circ \begin{bmatrix} 100 \\ 010 \\ 001 \end{bmatrix} = \begin{bmatrix} 001100 \\ 001010 \\ 001001 \end{bmatrix}$$

$$G^2 = S^1 \circ S^1 = \begin{bmatrix} 100 \\ 010 \\ 001 \end{bmatrix} \circ \begin{bmatrix} 100 \\ 010 \\ 001 \end{bmatrix} = \begin{bmatrix} 100100 \\ 010010 \\ 001001 \end{bmatrix}$$

The symbol "∘" denotes concatenation of matrices. The pixel expansion is $m^* = (g-1) \times m = 6$. The relative difference between the second-grey-level reconstructed pixel and the first-grey-level reconstructed pixel is $\alpha^{(1,0)} = 1/6$, and the relative difference between the third-grey-level reconstructed pixel and the second-grey-level reconstructed pixel is $\alpha^{(2,1)} = 1/6$. It is clear that $\alpha^{(1,0)} = \alpha^{(2,1)} = \alpha / (g-1) = 1/6$.

C. *Colored $(k, n, m', c)$-CVSS schemes*

The black–and–white $(k,n,m)$-VSS schemes have been extended to design a color $k$-out-of-$n$ visual secret sharing ($(k,n,m',c)$-CVSS) scheme for $c$ colors with pixel expansion $m'$ by Verheul and Van Tilborg [3]. The following definition is the formal definition of a colored $(k,n,m',c)$-CVSS schemes in [3] (also see [4] [5]).

***Definition 2***[3, 4, 5]: A $k$-out-of-$n$ $c$-color VSS scheme $S=(C_0,\cdots,C_{c-1})$, consists of $c$ collections of $n \times m'$ $q$-ary matrices, in which the $c$ colors are elements of the *Galois field* $GF(q)$. To share a pixel of color $i$, the dealer randomly chooses one of the matrices in $C_i$. The chosen matrix defines the color of the $m'$ sub-pixels in each one of the shares. The solution is considered valid if the following three conditions are met for all $0 \leq i \leq c-1$.

1. For any $S$ in $C_i$, the "OR" $V$ of any $k$ of the $n$ rows satisfies $z_i(V) \geq h$, where $V$ is a vector with coordinates in $c$ colors and black color, and $z_i(V)$ denotes the number of coordinates in $V$ equal to color $i$.

2. For any $S$ in $C_i$, the "OR" $V$ of any $k$ of the $n$ rows satisfies $z_j(V) \leq l$, for $j \neq i$.

3. For any $i_1 < i_2 < \cdots < i_s$ in $\{1,\cdots,n\}$ with $s<k$, the collections of $s \times m'$ matrix in $C_j$ to rows $i_1, i_2, \cdots, i_s$ are indistinguishable in the sense that they contain the same matrices with the same frequencies.

Note that $h>l$ and $m'$ is the block length of a colored VSS scheme. The relative difference $\alpha'_i = \dfrac{h-l}{m'}$ is for color image.

We illustrate the result of in [3, 5] with the following example.

***Example 3:*** Suppose the secret image has a color set of {red, green, blue} which is denoted as $\{r, g, b\}$. The basis matrices of a (2, 3, 7, 3)-CVSS scheme are [6]:

$$C_r = \begin{bmatrix} r\ g \bullet \bullet b \bullet \bullet \\ r \bullet g \bullet \bullet b \bullet \\ r \bullet \bullet g \bullet \bullet b \end{bmatrix}, \quad C_g = \begin{bmatrix} g\ r \bullet \bullet b \bullet \bullet \\ g \bullet r \bullet \bullet b \bullet \\ g \bullet \bullet r \bullet \bullet b \end{bmatrix}, \quad C_b = \begin{bmatrix} b\ g \bullet \bullet r \bullet \bullet \\ b \bullet g \bullet \bullet r \bullet \\ b \bullet \bullet g \bullet \bullet r \end{bmatrix}$$

The pixel expansion is $m' = 7$. And $h = 1$, $l = 0$, the relative difference $\alpha'_i = h - l / m' = 1/7$, $i = r, g, b$.

### D. Probabilistic VSS

To solve the pixel expansion problem with VSS schemes, Ito et al. [13], Yang [7] and Cimato [8] proposed probabilistic VSS models for black-and-white (binary) images. In reconstructing the secret image, the "OR" operation of pixels of the shadows is the same as the stacking operation of sub-pixels in the non-probabilistic VSS schemes. They defined $p_0$ (resp. $p_1$) as the appearance probability of white pixel in a white (resp. black) area of the recovered image. For a fixed threshold probability $0 \leq p_{TH} \leq 1$ and relative contrast $\alpha \geq 0$, if $p_0 \geq p_{TH}$ and $p_1 \leq p_{TH} - \alpha$, the frequency of white pixels in a white area of the recovered image should be higher than that in a black area. Yang proposed a probabilistic scheme for binary images with no pixel expansion, i.e, $m = 1$ [7]. Cimato [8] introduced probabilistic schemes with pixel expansion $m \geq 1$, generalizing the model of Yang [5] and proved that there is one-to-one correspondence between probabilistic schemes with no pixel expansion and deterministic schemes; such one-to-one mapping trades-in the probabilistic nature of the scheme with contrast of the deterministic scheme.

### E. Quality measures

Since the existing probabilistic schemes were only proposed for binary images, the contrast between black-and-white pixels was naturally chosen as an important quality measure. Our proposed scheme is for gray-scale images and color images. We use the expected contrast between two pixels with consecutive grey-levels in the original image to indicate the quality. We define it as "Average Contrast" in detail below.

***Definition 3:*** For any deterministic $(k, n, m^*, g)$-GVSS schemes, the basis matrices are $G^i$, $i = 0, \ldots, g-1$. Randomly selecting any $s$ columns from $G^i$ ($s = 1, \ldots, m^*$), we get $C^s_{m^*}$ $n \times s$ Boolean matrices $G^i|_{s,p}$ which consists of a set (collection) denoted as $T_s^i = \{G^i|_{s,p}\}$, $p = 1, \ldots, C_{m^*}^s$.

For any element $G^i|_{s,p}$ of the set $T_s^i$, $p_{s,j}^i$ denotes the probabilistic of the Hamming weight of the OR of any $k$ rows in the matrix is $j$, $j = 0, \ldots, s$. Thus, the average Hamming weight of the $i$-th grey-level reconstructed pixel is

$$\overline{H}_s^i = \sum_{j=0}^{s} j \cdot p_j^i \tag{1}$$

Average grey-levels is

$$\overline{\beta}_s^i = \frac{H_s^i}{s} \tag{2}$$

The Average relative difference between the $i$-th and the $(i+1)$-th the average grey-levels.

$$\overline{\alpha}^{i+1, i}|_s = \overline{\beta}_s^{i+1} - \overline{\beta}_s^i, \quad i = 0, \cdots, g-2 \tag{3}$$

Similarly, the following is the definition of a measurement for a probabilistic $(k, n, m', c)$-CVSS which is realized in a deterministic $(k, n, m', c)$-CVSS scheme.

***Definition 4:*** Suppose a $c$-colored deterministic $(k, n, m', c)$-CVSS scheme. The basis matrices are $C_i$, $i = 0, \ldots, c-1$. Selecting any $t$ columns from $C_i$ ($t = 1, \ldots, m'$), we get a set including $C_m^t$ $n \times t$ Boolean matrices $C_i|_{t,q}$, $q = 1, \ldots, C_{m'}^s$, denoted as $S_t^{(i)} = \{C_i|_{t,q}\}$. Assume $p_t^{(i|i)}$ is the probability of reconstructed color $i$ pixel for any matrix $C_i|_{t,q}$ in $S_t^{(i)}$. The probability of the OR of any $k$ rows of any matrix $C_i|_{t,q}$ including color $j$ is $p_t^{(j|i)}$.

Thus, the average Hamming weight of the reconstructed pixel with color $i$ is

$$\overline{H}_t^{(i|i)} = \sum_{j=0}^{t} j \cdot p_j^{(i|i)}, \quad i = 0, \cdots, c-1 \tag{4}$$

The average Hamming weight of the reconstructed pixel with color $j$ is

$$\overline{H}_t^{(j|i)} = \sum_{j=0}^{t} j \cdot p_j^{(j|i)} \tag{5}$$

The average contrast is of a constructed pixel is

$$\overline{\beta}_t^{(i|i)} = \frac{H_t^{(i|i)}}{t}, \quad \beta_t^{(j|i)} = \frac{H_t^{(j|i)}}{t} \tag{6}$$

The relative difference of the reconstructed secret image is

$$\overline{\alpha}_t^{(i,i)} = \overline{\beta}_t^{(i|i)} - \overline{\beta}_t^{(j|i)}, \quad i = 0, \cdots, c-1 \tag{7}$$

## III. THE PROPOSED $(k, n, m^*, t, g)$-PROB. GVSS SCHEME

Blundo et al. [2] and Muecke[4] constructed $g$ grey-level $(k, n, m^*, g)$-GVSS schemes using black-and-white $(k, n, m)$-VSS schemes. The matrices $S^0$ and $S^1$ are basis matrices of $(k, n, m)$-VSS schemes.

In the $(k, n, m^*, g)$-GVSS scheme with basic matrices $G^i$,

$$G^i = \underbrace{S^0 \circ \ldots \circ S^0}_{g-i-1} \underbrace{S^1 \circ \ldots S^1}_{i}, \quad i = 0, \cdots, g-1.$$

Suppose that $a_i$ is the number of ones in $G^i$. Let $b_i$ be the number of zeros in $G^i$, we have that

$$a_i = H(V^i), \quad b_i = m^* - a_i \tag{8}$$

Notice that when all the relative difference $\alpha_i$ ($i = 0, \cdots, g-2$) are equal, then the scheme proposed reduced to the one proposed by Naor and Shamir for *k*-out-of-*n* VSS [2]. Namely, the full gray-scale range from black to white and has fixed contrast difference between grey-levels. In this case,

$$a_i = (m-h) \cdot (g-i-1) + (m-l) \cdot i, \quad b_i = h \cdot (g-i-1) + l \cdot i \tag{9}$$

The pixel expansion is $a_i + b_i = m \times (g-1) = m^*$. The relative difference $\alpha_i$ between the *i*-th and (*i*+1)-th grey-level, $i = 0, \cdots, g-2$.

$$\alpha_i = \frac{(m-l) - (m-h)}{(g-1) \cdot m} = \frac{h-l}{(g-1) \cdot m} = \frac{\alpha}{g-1}, \quad i = 0, \cdots, g-2. \tag{10}$$

### A. The model of proposed probabilistic VSS for gray-scale image

Let $(k, n, m^*, g)$-GVSS be a deterministic scheme. The basic construction matrix for grey-level $i$ is $G^i$, here $i = 0, \cdots, g-1$. There are $C_{m^*}^s$ different ways of arbitrarily choosing s columns of the $m^*$ column vectors of $G^i$. These $C_{m^*}^s$ $n \times s$ matrices are denoted as $G^i\big|_{s, p}$, $p = 1, \ldots, C_{m^*}^s$. We shall write $T_s^i = \{G^i\big|_{s, p}\}$. This collection $T_s^i$, $i = 0, \cdots, g-1$, can be used to construct a gray-scale probabilistic GVSS scheme.

Suppose a $g$ grey-levels $(k, n, m^*, s, g)$-Prob. GVSS which represents a probabilistic VSSS based on a $g$ grey-levels deterministic $(k, n, m^*, g)$-GVSS Scheme with pixel expansion $s$, $s \in \{1, \cdots, m^*\}$. The following example demonstrates the model above.

*Example 4:  (continuation of Example 2)*

The basis matrices of a (2, 3, 6, 3)-VSS scheme with three grey-levels are shown $G^0$, $G^1$, $G^2$. Thus, in the (2, 3, 6, $s$, 3)-Prob. GVSS with three grey-levels, $s \in \{1, 2, 3, 4, 5, 6\}$. The following basic matrices are the special cases of $s$ with $s = 5$, and the other cases are similar.

When $s = 5$, the basic matrices of (2, 3, 6, 5, 3)-Prob. GVSS with three grey-levels are

$$T_5^{(0)} = \left\{ \begin{bmatrix} 01001 \\ 01001 \\ 01001 \end{bmatrix}, \begin{bmatrix} 01001 \\ 01001 \\ 01001 \end{bmatrix}, \begin{bmatrix} 00001 \\ 00001 \\ 00001 \end{bmatrix}, \begin{bmatrix} 00101 \\ 00101 \\ 00101 \end{bmatrix}, \begin{bmatrix} 00101 \\ 00101 \\ 00101 \end{bmatrix}, \begin{bmatrix} 00100 \\ 00100 \\ 00100 \end{bmatrix} \right\}$$

$$T_5^{(1)} = \left\{ \begin{bmatrix} 01100 \\ 01010 \\ 01001 \end{bmatrix}, \begin{bmatrix} 01100 \\ 01010 \\ 01001 \end{bmatrix}, \begin{bmatrix} 00100 \\ 00010 \\ 00001 \end{bmatrix}, \begin{bmatrix} 00100 \\ 00110 \\ 00101 \end{bmatrix}, \begin{bmatrix} 00110 \\ 00100 \\ 00101 \end{bmatrix}, \begin{bmatrix} 00110 \\ 00101 \\ 00100 \end{bmatrix} \right\}$$

$$T_5^{(2)} = \left\{ \begin{bmatrix} 00100 \\ 10010 \\ 01001 \end{bmatrix}, \begin{bmatrix} 10100 \\ 00010 \\ 01001 \end{bmatrix}, \begin{bmatrix} 10100 \\ 01010 \\ 00001 \end{bmatrix}, \begin{bmatrix} 10000 \\ 01010 \\ 00101 \end{bmatrix}, \begin{bmatrix} 10010 \\ 01000 \\ 00101 \end{bmatrix}, \begin{bmatrix} 10010 \\ 01001 \\ 00100 \end{bmatrix} \right\}$$

Before we obtain the following theorem 1, we need to present two lemmas. We denote by $\binom{a}{i}$ the number of $i$-combinations of an $a$-element set. Lemma 1 below is from Exercise 25 of Section 5.6 in [10].

**Lemma 1**[10]: For all positive integers $m_1$, $m_2$, and $m_3$,

$$\sum_{i=0}^{m_3}\binom{m_1}{i}\binom{m_2}{m_3-i}=\binom{m_1+m_2}{m_3}$$

**Lemma 2:** Let $i,a,b,t$ are positive integers, $t<\min\{a,b\}$. Then

$$\sum_{i=0}^{t} i\cdot\binom{a}{i}\binom{b}{t-i}=a\cdot\binom{a-1+b}{t-1}$$

**Proof:** Since $i\cdot\binom{a}{i}=a\cdot\binom{a-1}{i-1}$, thus

$$\sum_{i=0}^{t} i\cdot\binom{a}{i}\binom{b}{t-i}=\sum_{i=1}^{t}(i\cdot\binom{a}{i})\cdot\binom{b}{t-i}$$

$$=\sum_{i=1}^{t}(a\cdot\binom{a-1}{i-1})\cdot\binom{b}{t-i} \quad \text{let } i=i-1$$

$$=a\cdot\sum_{i=0}^{t-1}\binom{a-1}{i}\binom{b}{t-1-i} \quad \text{by Lemma 1}$$

$$=a\cdot\binom{a-1+b}{t-1}.$$

Now, we recall also that the set $T_s^i=\{G^i|_{s,p}\}$, $p=1,\ldots,C_{m^*}^s$, $s=1,\ldots,m^*$, the matrix $G^i|_{s,p}$ represents selecting any $s$ columns from basic matrix $G^i$. We now give the following theorem.

**Theorem 1:** For any deterministic $g$ grey-levels $(k,n,m^*,g)$-GVSS scheme with basis matrices $G^i$, i =1, …, g, there exists a probabilistic $g$ grey-levels $(k,n,m^*,s,g)$-Prob. GVSS scheme. The basis matrix collection $T_s^i$ includes $C_{m^*}^s$ $n\times s$ Boolean matrices $G^i|_{s,p}$. In the $g$ grey-levels $(k,n,m^*,s,g)$-Prob.VSS scheme, the average contrast of the *i-th* grey-level reconstructed pixel is

$$\overline{\beta}_s^i=\frac{H_s^i}{s}=\frac{a_i}{m^*}.$$

The relative difference between the *i-th* and the *i+1-th* grey-level reconstructed pixel is

$$\overline{\alpha}_s^{(i,i+1)} = \beta_s^{i+1} - \beta_s^i = \frac{a_{i+1} - a_i}{m^*}.$$

***Proof*:**

***Security condition*:**

In a deterministic $g$ grey-levels $(k, n, m^*, g)$-GVSS scheme, we know the fact that the Hamming weight of the OR of any $k'$, $k' < k$, rows is the same in the basis matrices $G^i$ in which $i = 1, \ldots, g$.

Selecting any $s$ columns from $G^i$, we get $C_{m^*}^s$ $n \times s$ Boolean matrices $T_{s,q}^{(i)}$ which consist of a set $T_s^{(i)} = \{T_{s,p}^{(i)}\}$, $p = 1, \ldots, C_{m^*}^s$. Obviously, the OR of any $k'$ rows in any matrix of $T_s^{(i)}$ in $T_s^{(i)}$ has the same Hamming weight, $i = 1, \ldots, g$. Therefore, the security condition is ensured.

***Contrast condition:***

In a probabilistic $g$ grey-levels $(k, n, m^*, s, g)$-Prob. GVSS scheme with a set $T_s^{(i)} = \{T_{s,p}^{(i)}\}$ in which $i = 1, \ldots, g$, $p = 1, \ldots, C_{m^*}^s$. We also know that $p_{s,j}^{(i)}$ is the probability of reconstructed color $j$, here $j = 0, \ldots, s$, it is easy to verify $\sum_{j=0}^{s} p_{t,j}^{(i)} = 1$. For convenience of further derivation, we use variables $a$ and $b$ to substitute $a_i$ and $b_i$. We obtain

$$p_{s,j}^{(i)} = \frac{\binom{a_i}{j}\binom{b_i}{s-j}}{\binom{m^*}{s}} = \frac{\binom{a}{j}\binom{b}{s-j}}{\binom{m^*}{s}} \tag{11}$$

From equation (1), (11), (2), we have

$$H_s^{(i)} = \sum_{j=0}^{s} j \cdot p_{s,j}^{(i)} = \frac{1}{\binom{m^*}{s}} \cdot \sum_{j=0}^{s} j \cdot \binom{a}{j}\binom{b}{s-j}$$

$$\beta_s^{(i)} = \frac{H_s^{(i)}}{s} = \frac{1}{s \cdot \binom{m^*}{s}} \cdot \sum_{j=0}^{s} j \cdot \binom{a}{j}\binom{b}{s-j} \quad \text{by **Lemma 2**}$$

$$= \frac{1}{s \cdot \binom{m^*}{s}} \cdot a \cdot \binom{a-1+b}{s-1}, \text{ by } m^* = a+b \quad (8)$$

$$= \frac{a}{m^*}$$

From equation (3), it is easy to obtain

$$\overline{\alpha}_s^{(i+1,i)} = \overline{\beta}_s^{(i+1)} - \overline{\beta}_s^{(i)} = \frac{a_i - a_{i-1}}{m^*} = \alpha_i, \; i = 0, \cdots, g-2$$

In a deterministic $(k, n, m(g-1), g)$-GVSS scheme, when we assume that the contrast (or relative difference) between two neighboring grey-levels is the same, we obtain the following corollary 1.

***Corollary 1:*** For any deterministic $g$ grey-levels $(k, n, m(g-1), g)$-GVSS, the basis matrices are $G^i$, $i = 0, \cdots, g-1$. If the difference of two neighboring grey-levels are the same, then there exists a probabilistic $g$ grey-levels $(k, n, m(g-1), s, g)$-Prob. GVSS scheme. The average contrast of the $i$-th grey-level reconstructed pixel

$$\overline{\beta}_s^{(i)} = \frac{\overline{H}_s^{(i)}}{s} = \frac{(m-h) \cdot (g-i) + (m-l) \cdot (i-1)}{(g-1) \cdot m}$$

The relative difference between the $i$-th and the $(i-1)$-th grey-level reconstructed pixels is

$$\overline{\alpha}_s^{(i,i-1)} = \overline{\beta}_s^{(i+1)} - \overline{\beta}_s^{(i)} = \frac{h-l}{(g-1) \cdot m} = \frac{\alpha}{g-1}, \text{ here } i = 0, \cdots, g-2.$$

Next, we illustrate the procedure above in the theorem 1 by an example of a (2, 3, 6, $s$, 3)-Prob. GVSS scheme with three grey-levels.

***Example 5:*** *(continuation of Example 4)*

The basis matrices of (2, 3, 6, 5, 3)-Prob. GVSS scheme with three grey-levels are $T_5^{(0)}, T_5^{(1)},$ and $T_5^{(2)}$. Here, the set $T_5^{(0)}$ includes six matrices in which the OR of the $q_1$-th row and the $q_2$-th row are 0, 1, 2, 3, 4, 5, and the number of the corresponding matrices are 0, 2, 4, 0, 0, 0, $q_1, q_2 \in \{1, 2, 3\}$ and $q_1 \neq q_2$. Thus, $p_{s,j}^{(i)}$ is

$p_{5,0}^{(0)} = p_{5,3}^{(0)} = p_{5,4}^{(0)} = p_{5,5}^{(0)} = 0/6 = 0$, $p_{5,1}^{(0)} = 2/6 = 1/3$, $p_{5,2}^{(0)} = 4/6 = 2/3$

From equation (2), $\overline{\beta}_s^{(i)} = \frac{1}{s} \cdot \sum_{j=0}^{s} j \cdot p_{s,j}^{(i)} = \frac{1}{5} \cdot (1 \cdot \frac{1}{3} + 2 \cdot \frac{2}{3}) = \frac{1}{3}$.

To use the similar method above, it is easy to get the following table II

TABLE II

VALUES OF AVERAGE GREY-LEVELS OF (2, 3, 6, 5, 3)-PROB. GVSS SCHEME

| $p_{5,j}^{(i)}$ | $j = 0$ | $j = 1$ | $j = 2$ | $j = 3$ | $j = 4$ | $j = 5$ | $\overline{\beta}_5^{(i)}$ |
|---|---|---|---|---|---|---|---|
| $i = 1$ | 0 | 1/3 | 2/3 | 0 | 0 | 0 | 1/3 |
| $i = 2$ | 0 | 0 | 1/2 | 1/2 | 0 | 0 | 1/2 |
| $i = 3$ | 0 | 0 | 0 | 2/3 | 1/3 | 0 | 2/3 |

The following table III lists values of $p_{s,j}^{(i)}$ and $\overline{\beta}_s^{(i)}$ in which $s = 1, \ldots, 6$, and $i = 1, 2, 3$.

TABLE III

VALUES OF AVERAGE GREY-LEVELS OF (2, 3, 6, $s$, 3)-PROB. GVSS SCHEME

For $i = 1$

|  | $j = 0$ | $j = 1$ | $j = 2$ | $j = 3$ | $j = 4$ | $j = 5$ | $j = 6$ | $\overline{\beta}_s^{(1)}$ |
|---|---|---|---|---|---|---|---|---|
| $p_{1,j}^{(1)}$ | 2/3 | 1/3 | - | - | - | - | - | 1/3 |
| $p_{2,j}^{(1)}$ | 6/15 | 8/15 | 1/15 | - | - | - | - | 1/3 |
| $p_{3,j}^{(1)}$ | 1/5 | 3/5 | 1/5 | 0 | - | - | - | 1/3 |
| $p_{4,j}^{(1)}$ | 1/15 | 8/15 | 6/15 | 0 | 0 | - | - | 1/3 |
| $p_{5,j}^{(1)}$ | 0 | 1/3 | 2/3 | 0 | 0 | 0 | - | 1/3 |
| $p_{6,j}^{(1)}$ | 0 | 0 | 1 | 0 | 0 | 0 | 0 | 1/3 |

For $i = 2$

|  | $j = 0$ | $j = 1$ | $j = 2$ | $j = 3$ | $j = 4$ | $j = 5$ | $j = 6$ | $\overline{\beta}_s^{(2)}$ |
|---|---|---|---|---|---|---|---|---|
| $p_{1,j}^{(2)}$ | 1/2 | 1/2 | - | - | - | - | - | 1/2 |
| $p_{2,j}^{(2)}$ | 3/15 | 9/15 | 3/15 | - | - | - | - | 1/2 |
| $p_{3,j}^{(2)}$ | 1/20 | 9/20 | 9/20 | 1/20 | - | - | - | 1/2 |
| $p_{4,j}^{(2)}$ | 0 | 1/5 | 3/5 | 1/5 | 0 | - | - | 1/2 |
| $p_{5,j}^{(2)}$ | 0 | 0 | 1/2 | 1/2 | 0 | 0 | - | 1/2 |
| $p_{6,j}^{(2)}$ | 0 | 0 | 0 | 1 | 1 | 0 | 0 | 1/2 |

For $i = 3$

| | $j = 0$ | $j = 1$ | $j = 2$ | $j = 3$ | $j = 4$ | $j = 5$ | $j = 6$ | $\overline{\beta}_s^{(3)}$ |
|---|---|---|---|---|---|---|---|---|
| $p_{1,j}^{(3)}$ | 1/3 | 2/3 | - | - | - | - | - | 2/3 |
| $p_{2,j}^{(3)}$ | 1/15 | 8/15 | 6/15 | - | - | - | - | 2/3 |
| $p_{3,j}^{(3)}$ | 0 | 1/5 | 3/5 | 1/5 | - | - | - | 2/3 |
| $p_{4,j}^{(3)}$ | 0 | 0 | 6/15 | 8/15 | 1/15 | - | - | 2/3 |
| $p_{5,j}^{(3)}$ | 0 | 0 | 0 | 2/3 | 1/3 | 0 | - | 2/3 |
| $p_{6,j}^{(3)}$ | 0 | 0 | 0 | 0 | 0 | 0 | 0 | 2/3 |

From the Table III above, it is clear that for a fixed value $i \in \{1, 2, 3\}$, the values of $\overline{\beta}_s^{(i)}$ are equal, while $s \in \{1,\cdots,6\}$. In other words, it verifies the results of Theorem 1 and Corollary 1).

### B. g grey-level (n, n, m*, s, g)-Prob. GVSS schemes

In a binary $(n, n)$-VSS[1], pixel expansion is $m = 2^{n-1}$, the parameters $h=1$, $l=0$. If we degenerate the above $g$ grey-levels $(k,n,m^*,g)$-GVSS to a $g$ grey-levels $(n, n, 2^{n-1}, g)$-GVSS schemes, we obtain the following the result is from the **Theorem 1**.

***Corollary 2:*** For any deterministic $g$ grey-levels $(n,n,m^*,g)$-GVSS, $m^* = m \cdot (g-1)$, the basis matrices are $G^i$, $i = 1, \ldots, g$. Thus, there exists a probabilistic $g$ grey-levels $(n, n, m^*, s, g)$-Prob. GVSS, the average contrast for the $i$-th grey-level is

$$\overline{\beta}_s^{(i)} = \frac{\overline{H}_s^{(i)}}{s} = \frac{(m-h)\cdot(g-i-1)+(m-l)\cdot i}{(g-1)\cdot m}, \text{ here } m = 2^{n-1}$$

The relative difference between the $i$-th and the $i+1$-th grey-levels is

$$\overline{\alpha}_s^{(i+1,i)} = \overline{\beta}_s^{(i)} - \overline{\beta}_s^{(i-1)} = \frac{h-l}{(g-1)\cdot 2^{n-1}}, i = 0,\cdots,g-2.$$

$$\overline{\alpha}_s^{(i+1,i)} = \frac{1}{(g-1)\cdot 2^{n-1}}, \text{ while } h = 1, l = 0.$$

### C. Binary (k, n, m, s)-Prob. VSS schemes

In sub-section A above, we construct a $g$ grey-levels $(k, n, m^*, s)$-Prob. VSS. When $g = 2$, it can degenerate to a binary $(k, n, m, s)$-Prob. VSS scheme. From **Theorem 1**, we get the following corollary 3.

***Corollary 3:*** In a binary $(k, n, m, s)$-Prob. VSS scheme, $s = 1, \ldots, m$, the average contrast of the reconstructed black pixel and white pixel are

$$\overline{\beta}_s^{(0)} = \frac{\overline{H}_s^{(0)}}{s} = \frac{m-h}{m}$$

$$\overline{\beta}_s^{(1)} = \frac{\overline{H}_s^{(1)}}{s} = \frac{m-l}{m}$$

The relative difference is

$$\overline{\alpha}_s^{(1,0)} = \overline{\beta}_s^{(1)} - \overline{\beta}_s^{(0)} = \frac{h-l}{m}$$

Obviously, when $s = 1$, this scheme is equivalent to [7] binary $(k, n, m, 1)$-Prob.VSS schemes.

## IV. The recognition area of a $(k, n, m^*, s, g)$-Prob. GVSS scheme

In a deterministic binary VSS scheme, there exists the difference between a reconstructed black pixel and a reconstructed white pixel. But in a probabilistic VSS schemes, the difference is not maintained. The average Hamming weight of a small black area is different from that of a white area. These small areas can be recognized as black or white correctly. Yang discussed the relationship between black small area and white small area in a binary $(k, n, 1)$-Prob. VSS schemes [14]. We will soon discuss the problem for our proposed Prob. GVSS scheme.

In a $g$ grey-levels $(k, n, m^*, s, g)$-Prob. GVSS scheme, assume that random variable $X_i$ represents the number of 1's in the $i$-th grey-level of the reconstructed secret image, thus the probability of $X_i = j$ is

$$p_{s,j}^{(i)} = P(X_i = j) = \binom{b_i}{s-j}\binom{a_i}{j} / \binom{m^*}{s}, \text{ where } \max(0, s-b_i) \leq j \leq \min(s, a_i).$$

Note that $p_{s,j}^{(i)}$ has a Hyper-geometric distribution with the following mean and variance (see [12]).

$$E(X_i) = s \cdot \frac{a_i}{m^*}, \text{Var}(X_i) = \frac{s \cdot (m^*-s) \cdot a_i \cdot (m^*-a_i)}{(m^*-1) \cdot (m^*)^2}$$

Let $N$ independent random variables $X_i^{(1)}, \ldots, X_i^{(N)}$ represent the number of 1's in the $i$-th grey-level reconstructed pixel. Thus, the mean and variance of the Hyper-geometric

$$SN_i = \sum_{i=1}^{N} X_i \text{ are}$$

$$\mu_i = s \cdot N \cdot \frac{a_i}{m^*}, \quad \sigma_i^2 = \frac{s \cdot N \cdot (m^*-s) \cdot a_i \cdot (m^*-a_i)}{(m^*-1) \cdot (m^*)^2}$$

Example 6 below gives the distribution curve of $(2, 3, 6, s, 3)$-Prob. VSS scheme.

***Example 6:*** *(continuation of Example 2)* The basis matrices of a deterministic (2, 3, 6, 3)-VSS Scheme with three grey-levels are $G^0$, $G^1$, and $G^2$.

The *i-th* grey-levels includes $a_i$ 1's and $b_i$ 0's

$$a_1=4, \ a_2=5, \ a_3=6$$

$$b_1=2, \ b_2=1, \ b_3=0$$

Suppose $N = 100$. Thus, the mean and the variance of the number of 1's in the reconstructed pixel with the *i-th* grey-leve for several values of *i* are given below.

For $i = 1$

| s | 1 | 2 | 3 | 4 | 5 | 6 |
|---|---|---|---|---|---|---|
| $\mu$ | 66.7 | 133.3 | 200 | 266.7 | 333.3 | 400 |
| $\sigma^2$ | 22.2 | 35.6 | 40.0 | 35.6 | 22.2 | 0 |

For $i = 2$

| s | 1 | 2 | 3 | 4 | 5 | 6 |
|---|---|---|---|---|---|---|
| $\mu$ | 83.3 | 166.7 | 250 | 333.3 | 416.7 | 500 |
| $\sigma^2$ | 13.9 | 22.2 | 25 | 22.2 | 13.9 | 0 |

For $i = 3$

| s | 1 | 2 | 3 | 4 | 5 | 6 |
|---|---|---|---|---|---|---|
| $\mu$ | 100 | 200 | 300 | 400 | 500 | 600 |
| $\sigma^2$ | 0 | 0 | 0 | 0 | 0 | 0 |

The following is the distribution curves of **Example 6** on 3000 random data sets. Each data set contains 100 random numbers.

The *X*-axes represents the value of $SN_i = \sum_{i=1}^{N} X_i$, the *Y*-axes represents the numbers of $SN_i$.

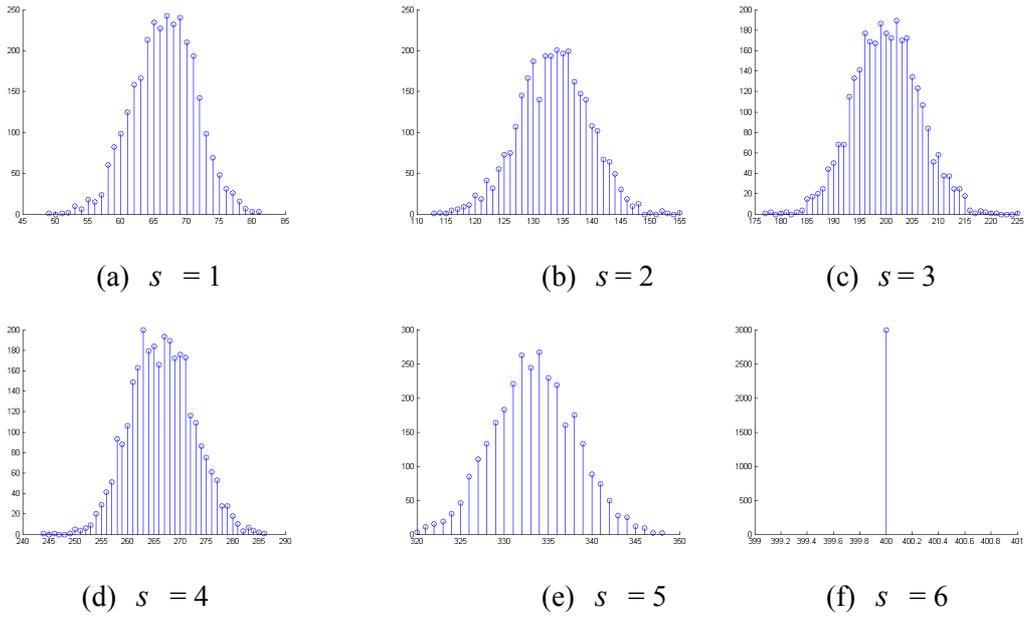

Fig. 1. The distribution curves of the reconstructed pixel with the first grey-level.

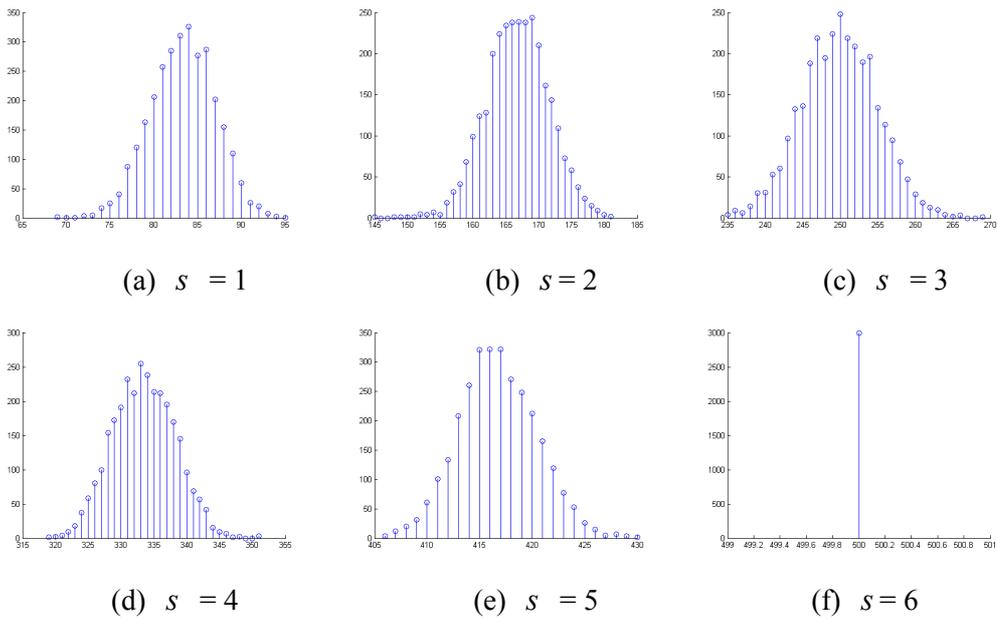

Fig. 2. The distribution curves of the reconstructed pixel with the second grey-level.

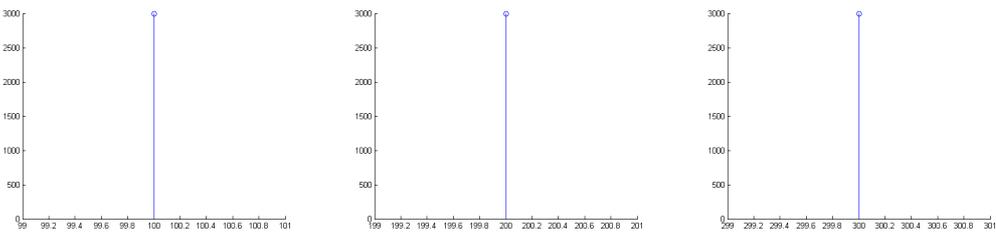

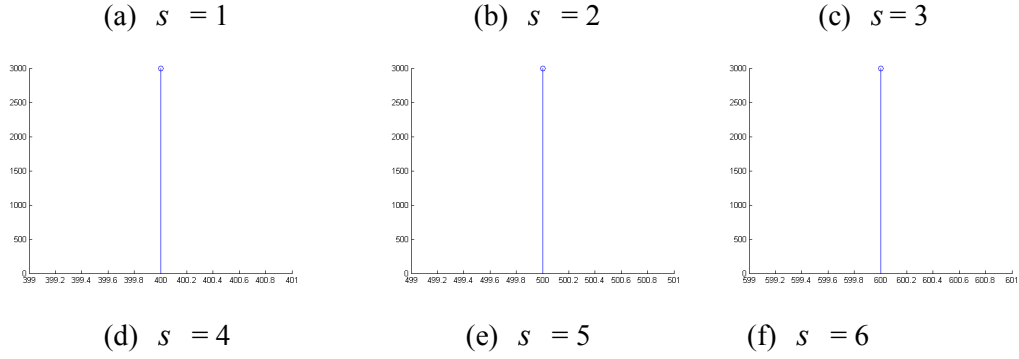

(a) $s = 1$  (b) $s = 2$  (c) $s = 3$

(d) $s = 4$  (e) $s = 5$  (f) $s = 6$

Fig. 3.  The distribution curves of the reconstructed pixel with the third grey-level.

The Hyper-geometric distribution curves are shown in Fig. 2, 3 and 4, and the solid lines are their normal approximations.

In a $g$ grey-levels $(k, n, m^*, g)$-GVSS scheme, $m^*$ is in the range of hundreds of thousands. For example, when $g = 256$, $m^* = 510$ in a (2, 2, 256) scheme and $m^* = 2040$ in a (4, 4, 256) scheme.

In a Hyper-geometric distribution, if $m^*$ goes to infinity, we have

$$\underset{m^* \to \infty}{Lim} \frac{a_i}{m^*} = \underset{m \to \infty}{Lim} \frac{(m-h)(g-i)+(m-l)(i-1)}{(g-1) \times m} = 1$$

$$m^* = (g-1) \times m, \; a_i = (m-h).(g-i)+(m-l)(i-1)$$

Based on Central Limit Theory, these Hyper-geometric distribution functions can be approximated by Normal distribution. And with a normal distribution, there are 99.73% data in $\mu \pm 3 \cdot \sigma$ area. So we use the following criterion to measure the recognition area

$$\mu_i - 3 \cdot \sigma_i > \mu_{i+1} + 3 \cdot \sigma_{i+1} + N^{(i+1, i)} \cdot d_{i+1}$$

$$0 < d_{i+1} < p_i - p_{i+1}$$

In a grey-level $(k, n, m^*, t, g)$-Prob. GVSS, the recognition area between the $i+1$-th and the $i$-th reconstructed pixels are

Let $d_{i+1} = d$

$$N^{(i+1, i)} > 9 \cdot \frac{s \cdot (m^*-s)}{m^*-1} \cdot \left( \frac{\sqrt{a_{i+1} \cdot (m^*-a_{i+1})} + \sqrt{a_i \cdot (m^*-a_i)}}{s \times a_i - s \times a_{i+1} - m^* \cdot d} \right)^2 \quad (12)$$

In (12), when $s = m^*$, $N^{(i+1, i)} = 0$, it is a deterministic scheme. When $s = 1$, if $g = 2$ （binary scheme）, the recognition area is

$$N^{(1,0)} > 9 \cdot \left( \frac{\sqrt{p_0 \times (1-p_0)} + \sqrt{p_1 \times (1-p_1)}}{p_0 - p_1 - d} \right)^2$$

In Yang's scheme [7], $p_0 = \frac{h}{m}$, $p_1 = \frac{l}{m}$. In our scheme $p_0 = \frac{m-h}{m}$, $p_1 = \frac{m-l}{m}$. It is easy to verify that the values of $N^{(1,0)}$ are equivalent to the size of the recognized area of Yang's Prob. VSS scheme when $p_0 = \frac{h}{m}$, $p_1 = \frac{l}{m}$ or $p_0 = \frac{m-h}{m}$, $p_1 = \frac{m-l}{m}$.

## V. PROPOSED COLORED PROB. VSS SCHME

In this section we will construct colored visual threshold probabilistic schemes.

In [3], a color image is seen as an array of pixels, each of which is $k_0, k_1, \ldots, k_{c-1}$. Here $c$ is the number of colors and $k_i$ is called the *i-th* color. Clearly, the grey-levels of a gray-scale image can be viewed as different colors. Each pixel is divided into $m'$ sub-pixels of color 0, ..., *c*-1.

When sub-pixels are put on top of each other and held to the light, one sees a "generalized" or, i.e. if all sub-pixels are of color *i* then one sees light of color *i*, otherwise one sees no light at all (i.e. black). The symbol ● represents black, the sign ● is always distinguishable from the *c* colors.

The "OR" operation of the two sub-pixels in Verheul and Van Tilborg scheme [3] is shown as following:

$$\text{color } i + \text{color } i = \text{color } i$$

$$\text{color } i + \text{color } j = \bullet, \text{ if } i \neq j$$

$$\text{color } i + \bullet = \bullet, \text{ here } i = 0, \ldots, c\text{-}1$$

The generalized "or" of elements in $\{k_0, k_1, \ldots, k_{c-1}\}$ equals $k_i$ of all elements are equal to $k_i$, otherwise it equals ● .For a vector $V$ with in $\{k_0, k_1, \ldots, k_{c-1}\} \cup (\bullet)$. Let $z_i(V)$ ($i = 0, \cdots, c-1$) denote the number of coordinates in $V$ equal to color $i$. In [3], the color $i$ circle sub-pixel with a sector or color $i$ and the other sector of black color cannot be directly used in the image editing package.

Yang used block to represent circle in [3] and redefine "OR" operation as followings [5].

$$\text{color } i + \text{color } i = \text{color } i$$

color $i$ + color $j$ =No definition when $i \neq j$;

color $i$ + ● = ●, here $i = 0,…, c$-1.

We now can determine some parameter values of the deterministic colored $(k,n,m',c)$-CVSS scheme according to sub-pixels operation methods above.

Suppose the original secret image has $c$ colors, and the color $i$ reconstructed pixel includes $a_i$ black sub-pixels, $b_i$ color $i$ sub-pixels, and $e_j$ color j sub-pixels.

$$m' = a_i + b_i + e_j \tag{13}$$

In **Definition 2**, $h \geq 1$, $l = 0$.

In (13), $e_j = 0$, $b_i = h$, $m' = a_i + b_i + e_j = a_i + b_i$, thus $\overline{H}_t^{(j|i)} = 0$, $j \neq i$.

Theorem 2 below is form results of Ref. [3].

***Theorem 2:*** Consider a deterministic colored $(k,n,m',c)$-CVSS with its relative difference of the reconstructed color $i$ being $\alpha_i'$, $i = 0,\cdots,c-1$. There exists a colored $(k,n,m',t,c)$-Prob. CVSS with pixel expansion $t$, $t \in \{1, ..., m'\}$,

The average contrast of the reconstructed color $i$ is

$$\overline{\beta}_t^{(i|i)} = \frac{\overline{H}_t^{(i|i)}}{t} = \frac{b_i}{m'} = \frac{h}{m'}, \quad \overline{\beta}_t^{(j|i)} = \frac{\overline{H}_t^{(j|i)}}{t} = 0$$

The average relative difference of the reconstructed color $i$ is

$$\overline{\alpha}_t^{(i,i)} = \overline{\beta}_t^{(i|i)} - \overline{\beta}_t^{(j|i)} = \frac{b_i}{m'} = \frac{h}{m'} \alpha_i'.$$

***Proof:***

The proof of **Theorem 2** is similar to the proof of **Theorem 1**. We only give a few highlighted posints here.

In a deterministic $c$ colored $(k,n,m',c)$-CVSS scheme, $a_i + b_i = m'$. In the corresponding $(k,n,m',s,c)$-Prob. CVSS scheme, suppose random variable $X_i$ represents the number of color $i$ of the reconstructed color $i$ pixel $U_s$. $X_i$ is equivalent to selecting $s$ sub-pixels from the reconstructed pixel $U$ in a deterministic CVSS scheme.

The probability of the color $i$ reconstructed pixel including $z$ color $i$ sub-pixels is

$$p_{t,z}^{(i|i)} = \frac{C_{a_i}^{t-z} C_{b_i}^{z}}{C_{m'}^{t}}, \quad z = 0, …, s.$$

The average number of the color $i$ sub-pixels in the color $i$ reconstructed pixels is

$$\overline{H}_t^{(i|i)} = \sum_{z=0}^{t} z \cdot p_{t,z}^{(i|i)}$$

From equivalent (2, 3, 7), we are enough to prove the following the result.

$$\alpha_t^{(i,i)} = \beta_t^{(i|i)} - \beta_t^{(j|i)} = \frac{b_i}{m'} = \frac{h}{m'}.$$

We now review the optimal deterministic $c$-color $(n, n)$-VSS scheme in [13, 15] and describe its relation to our proposed color scheme.

***Lemma 3[15]:*** The pixel expansion of a $c$-color $(n, n)$-VSS scheme, for any $c$, $n \geq 2$, is lower bounded by

$$m' \geq \begin{cases} c \cdot 2^{n-1} - 1, & \text{if } n \text{ is even} \\ c \cdot 2^{n-1} - c + 1, & \text{if } n \text{ is odd} \end{cases}$$

The optimal contrast of a $c$-color $(n, n)$-VSS scheme is

$$\alpha'_{opt} \geq \begin{cases} \dfrac{1}{c \cdot 2^{n-1} - 1}, & \text{if } n \text{ is even} \\ \dfrac{1}{c \cdot 2^{n-1} - c + 1}, & \text{if } n \text{ is odd} \end{cases}$$

The following corollary is directly from the Theorem 2 and Lemma 3.

***Corollary 4:*** Suppose that in a deterministic optimal colored $(n, n, m', c)$-CVSS scheme the relative difference of the reconstructed color $i$ is $\alpha'_{opt}$, $i = 0, \cdots, c-1$. Thus, there exists a colored $(n, n, m', t, c)$-Prob. CVSS with pixel expansion $t$, $t \in \{1, ..., m'\}$, and the average relative difference of the reconstructed color $i$ is

$$\overline{\alpha}_t^{(i,i)} \geq \begin{cases} \dfrac{1}{c \cdot 2^{n-1} - 1}, & \text{if } n \text{ is even} \\ \dfrac{1}{c \cdot 2^{n-1} - c + 1}, & \text{if } n \text{ is odd} \end{cases} = \alpha'_{opt}.$$

## VI. THE RECOGNITION AREA OF A COLORED $(k, n, m', t, c)$-PROB. CVSS SCHEME

In this section, we will discuss the recognition area of a colored $(k, n, m', t, c)$-Prob.

CVSS scheme. Although **Definition 2** gave a general model to describe color VSS scheme, the pixel expansion and relative difference in some proposed colored $(k, n, m', c)$-VSS schemes are limited by the condition of $l = 0$ and $h \geq 1$ [ 3, 5, 13, 15]. We discuss the size of recognition area under this limited condition.

In a colored $(k, n, m', c)$-VSS scheme, color $i$ reconstructed pixel includes $h$ color $i$ sub-pixels and $l$ color $j$ sub-pixels, $i \neq j$. Suppose the original secret image has $c$ colors, color $i$ reconstructed pixel includes $a_i$ black pixels and $b_i$ color $i$ sub-pixels:

$$a_i = m' - b_i$$

Next, we analyze the recognition area of a colored $(k, n, m', t, c)$-Prob. CVSS scheme. Assume the random variable $X_i$ represents the number of color $z$ sub-pixels in a color $i$ reconstructed pixel. Thus, the probability of $X_i = z$ is

$$p_{t,z}^{(i|i)} = \frac{C_{a_i}^{t-z} C_{b_i}^{z}}{C_{m'}^{t}}$$

and the mean and variance of the Hyper-geometric $X_i$ are

$$E(X_i) = t \cdot \frac{b_i}{m'}, \quad Var(X_i) = \frac{t \cdot (m'-t) \cdot b_i \cdot (m'-b_i)}{(m'-1) \cdot (m')^2}. \tag{14}$$

Consider the $N$ random variables $X_i^{(1)}, \ldots, X_i^{(N)}$ representing the number of color $i$ sub-pixels in the color $i$ reconstructed pixel. The mean and variance of the Hyper-geometric $SN_i = \sum_{i=1}^{N} X_i$ are

$$\mu_i = N \cdot t \cdot \frac{b_i}{m'} \tag{15}$$

$$\sigma_i^2 = N \cdot \frac{t \cdot (m'-t) \cdot b_i \cdot (m'-b_i)}{(m'-1) \cdot (m')^2} \tag{16}$$

*Example 7:   (continuation of Example 3)*

Let the color set of the secret image be $\{r, g, b\}$. The basis matrices of a deterministic colored (2, 3, 7, 3) –VSS Scheme are $C_r$, $C_g$, and $C_b$

In color $i$ reconstructed pixel, the number $a_i$ of black sub-pixels and the number $b_i$ of color $i$ sub-pixels are

$$a_r = 6, \; a_g = 6, \; a_b = 6$$

$$b_r = 1, \; b_g = 1, \; b_b = 1.$$

When $N$ = 100, the mean and variance of the color $c$ sub-pixels in a color $c$

reconstructed pixel are

| T | 1 | 2 | 3 | 4 | 5 | 6 | 7 |
|---|---|---|---|---|---|---|---|
| $\mu$ | 14 | 29 | 43 | 57 | 71 | 86 | 100 |
| $\sigma^2$ | 12.2 | 20.4 | 24.5 | 24.5 | 20.4 | 12.2 | 0 |

The following is the distribution curves of **Example 7**.

The $X$-axis represents the value of $SN_i = \sum_{i=1}^{N} X_i$, the $Y$-axis represents the numbers of $SN_i$.

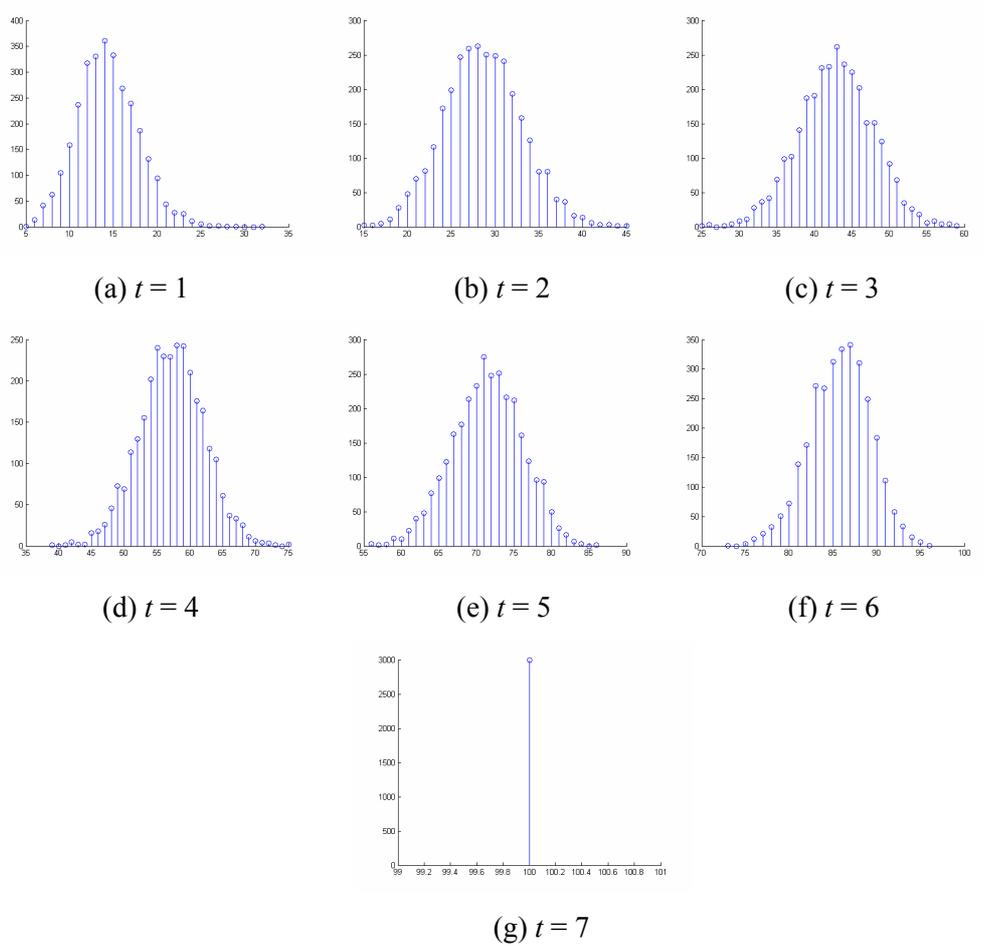

(a) $t = 1$    (b) $t = 2$    (c) $t = 3$

(d) $t = 4$    (e) $t = 5$    (f) $t = 6$

(g) $t = 7$

Fig. 5.  The distribution curves of the reconstructed pixel with color $i$.

The curves in Fig. 5 show the Hyper-geometric distributions and the solid lines are their normal approximations. In a colored VSS scheme, $m'$ is in the range of hundreds of thousands. Based on Central Limit Theory, these Hyper-geometric distribution functions can be safely approximated by Normal distribution. Therefore, the criterion of the measurement of recognition area is

$$\mu_{i|i} - 3 \cdot \sigma_{i|i} > \mu_{j|i} + 3 \cdot \sigma_{j|i} + N \cdot d_i, \quad 0 < d_i < p_t^{(i|i)} - p_t^{(j|i)}.$$

And, $\mu_{j|i}=0$ and $\sigma_{j|i}=0$. So we have

$$\mu_{i|i} - 3 \cdot \sigma_{i|i} > N \times d_i.$$

For convenience, let $N^{(i,i)}$ denote the recognition area. Here $N^{(i,i)} = N$. From (13), (15),

$$u_{i|i} = \mu_i = N^{(i,i)} \cdot t \cdot \frac{b_i}{m'}, \quad \sigma_{i|i} = \sigma_i$$

$$\sigma_{i|i}^2 = N^{(i,i)} \cdot \frac{t \cdot (m'-t) \cdot b_i \cdot (m'-b_i)}{(m'-1) \cdot (m')^2},$$

$$0 < d_i < p_t^{(i|i)} - p_t^{(j|i)}$$

$$\mu_{i|i} - 3\sigma_{i|i} > N^{(i,i)} \cdot d_i$$

In a colored $(k, n, m', t, c)$-Prob.VSS scheme, the recognition area of a color $i$ reconstructed pixel is

$$N^{(i,i)} > 9 \frac{t(m'-t) \cdot b_i \cdot (m'-b_i)}{(m'-1)(t\, b_i - m' d_i)^2} \tag{17}$$

When $b_i = 1$, (17) becomes

$$N^{(i,i)} > 9 \cdot \frac{t \cdot (m'-t)}{(t - m' \cdot d_i)^2} \tag{18}$$

When $t=1$ and $b_i=1$, it is a binary Prob. VSS scheme, we obtain

$$N^{(i,i)} > 9 \cdot \frac{(m-1)}{(1 - m \cdot d_i)^2}$$

In Yang's scheme, $N^{(i,i)} > 9 \cdot \left(\frac{\sqrt{p_o(1-p_0)} + \sqrt{p_1(1-p_1)}}{p_0 - p_1 - d}\right)^2$.

If $p_0 = 1/m$, $p_1 = 0$ then $N^{(i,i)} > 9 \cdot \frac{(m-1)}{(1 - m \cdot d_i)^2}$, Yang's scheme is a special case of (18). When $t = m$, $N^{(i,i)} = 0$, our proposed Prob. color CVSS scheme is a deterministic color CVSS scheme.

## VII. CONCLUSION

We have proposed two new probabilistic schemes for gray-scale images and color images. The pixel expansion of our schemes is from 1 (no pixel expansion) to a user-specified value. The quality of the reconstructed image, measured in terms of relative difference is the same as the conventional deterministic VSS schemes. Our technique can used to extend almost any existing deterministic VSS schemes to Prob. VSS schemes. Some deterministic color CVSS schemes in [16, 17] discussed the case in which the parameter $l \neq 0$, we are currently investigating the approaches of extending these schemes to a more general probabilistic color VSS scheme when $l \neq 0$.

## REFERENCES

[1] M. Naor and A. Shamir, "Visual cryptography," *advances in cryptology-EUROCRYPT'94, Le ture Notes in Computer Science*, vol. 950, pp. 1-12, 1995.

[2] C. Blundo, A. De Santis, and M Naor, "Visual cryptography for grey level images," *Information Processing Letters*, vol. 27, pp. 255-259, 2000.

[3] E. R. Verheul and H. C. A. Van Tilborg, "Constructions and Properties of k-out-of-n visual secret sharing schemes," *Designs, Codes and Cryptography*, vol.11, pp.179-196, 1997.

[4] Innes Muecke, "Greyscale and colour visual cryptography," thesis of the degree of Master of computer science, *Dalhouse University –Daltech*, 1999.

[5] C. N. Yang, C. S. Laih, "New colored visual secret sharing schemes," *Designs, Codes and Cryptography*, vol. 20, pp. 325-335, 2000.

[6] R. Ito, H. Kuwakado, and H. Tanaka, "Image size invariant visual cryptography," *IEICE Trans. Fundamentals*, vol. E82-A, no.10, pp. 2172-2177, 1999.

[7] C. N. Yang, "New visual secret sharing schemes using probabilistic method," *Pattern Recognition Letters*, vol. 25, pp. 481-494, 2004.

[8] S. Cimato, R. De Prisco, and A. De Santis, "Probabilistic visual cryptography schemes," *The computer Journal*, vol. 49, no. 1, pp. 97-107, 2006.

[9] C. S. Hsu, S. F, Tu and Y. C. Hou, "An optimization model for visual cryptography schemes with unexpanded shares," Foundations of Intelligent Systems-16th International Symposium, *Lecture Notes in Computer Science*, vol. 4203, pp. 58-67, 2006.

[10] R. A. Brualdi, "*Introductory Combinatorics* (Third Edition)," *Prentice Hall*, 1999.

[11] M. Iwamoto and H. Yamamoto, "The optimal *n*-out-of-*n* visual secret sharing scheme

for gray-scale images," *IEICE Trans. Fundamentals*, vol. E85-A, no.10, pp. 2238-2247, 2002.

[12] J. L. Devore, "Probability and Statistics: for Engineering and The Sciences," *Brooks/Cole, a division of Thomson Learing*, 2000.

[13] S. Cimato, R. De Prisco, and A. De Santis, "Optimal colored threshold visual cryptography schemes," *Designs, Codes and Cryptography*, vol. 35, no.3, pp.311-335, 2005.

[14] C. N. Yang and T. S. Chen, "Visual secret sharing scheme: improving the contrast of a recovered image via different pixel expansion," Image Analysis and Recognition, *Lecture Notes in Computer Science*, vol. 4141, pp. 468-479, 2006.

[15] S. Cimato, R. De Prisco and A. De Santis, "contrast optimal colored visual cryptography schemes," *IEEE Information Theory Workshop*, pp.139-142, 2003.

[16] V. Rijmen and B. Preneel, "Efficient colour visual encryption or 'Shared Colors of Benetton'," Rump Session of EUROCRYPTO'96, 1996. Available as http://www.iacr.org/conferences/ec96/rump/preneel.ps

[17] S. J. Shyu, "Efficient visual secret sharing scheme for color images," *Pattern Recognition*, vol. 39, no.5, pp. 866-880, 2006.

# APPENDIX A.   NOTATION

| | |
|---|---|
| Black-and-white $(k,n,m)$-VSS Scheme | Basic Boolean matrices $n \times m$   $S^0, S^1$ <br><br> Pixel expansion $m$, <br><br> Relative difference $\alpha = (h-l)/m$ |
| $g$ Grey-level image $(k,n,m^*,g)$-GVSS Scheme | Basic matrices $G^i$, $i = 0, \cdots, g-1$ <br><br> Pixel expansion $m^*$ <br><br> Let $a_i$ Be the number of ones in $G^i$, $b_i$ be the number of zeros in $G^i$.  $a_i = H(V^i)$, and $b_i = m^* - a_i$ <br><br> The relative difference $\alpha_i$ between the $i$ th and $(i+1)$th grey-levels, $i = 0, \cdots, g-2$. |
| $c$ Colored image $(k,n,m',c)$-CVSS Scheme | Basic matrices $C_i$, $i = 0, \cdots, c-1$, <br> Pixel expansion $m'$, <br> Relative differences, $\alpha'_i, i = 0, \cdots, c-1$. |
| $g$ Grey-level image Prob. $(k,n,m^*,s,g)$ VSS Scheme | Pixel expansion $s$  ($s=1,\ldots, m^*$) <br> Basic matrices $T_s^i = \{G^i\big|_{s,p}\}$,  $p = 1, \ldots, C_{m^*}^s$ <br> Average grey-levels $\overline{\beta}_s^i = \overline{H}_s^i / s$ <br> The Average relative difference $\overline{\alpha}^{i+1,i}\big|_s = \overline{\beta}_s^{i+1} - \overline{\beta}_s^i$ <br> $i = 0, \cdots, g-2$. |
| $c$ colored image Prob. $(k,n,m',s,c)$-CVSS Scheme | Pixel expansion $t$  ($t=1,\ldots, m'$) <br> Basic matrices $S_t^i = \{C_i\big|_{t,q}\}$,  $q = 1, \ldots, C_{m'}^t$ <br> Average grey-levels $\overline{\beta}_t^{(i|i)} = \overline{H}_t^{(i|i)} / t$ <br> The Average relative difference $\overline{\alpha}_t^{(i,i)} = \overline{\beta}_t^{(i|i)} - \overline{\beta}_t^{(j|i)}$ <br> $i,j = 0, \cdots, c-1$,  $j \neq i$ |